\documentclass{iaus}
\usepackage{graphicx}
\usepackage{natbib}
\usepackage{amsmath}

\newcommand{\Ms}{\ensuremath{M_\odot}}

\def\araa{ARA\&A}%
\title[The ages of Galactic globular clusters] 
{The ages of Galactic globular clusters\\ in the context of self-enrichment}

\author[Decressin et al.]   
{T. Decressin$^1$,
  H. Baumgardt$^1$,
  P. Kroupa$^1$,
  G. Meynet$^2$
 \and C. Charbonnel$^{2,3}$
}

\affiliation{$^1$Argelander Institute for Astronomy (AIfA), Auf dem
                H\"ugel 71, D-53121 Bonn, Germany \\ 
                email: {\tt decressin@astro.uni-bonn.de,
                  holger@astro.uni-bonn.de, pavel@astro.uni-bonn.de,} \\[\affilskip]
$^2$Geneva Observatory, University of Geneva, chemin des
     Maillettes 51, CH-1290 Sauverny, Switzerland \\
email: {\tt Georges.Meynet@unige.ch, Corinne.Charbonnel@unige.ch} \\[\affilskip]
$^3$LATT, CNRS UMR 5572, Universit\'e de Toulouse, 14 avenue Edouard
               Belin, F-31400 Toulouse Cedex 04, France
}

\pubyear{2008}
\volume{IAU258}  
\pagerange{119--126}
\jname{Title of your IAU Symposium}
\editors{A.C. Editor, B.D. Editor \& C.E. Editor, eds.}
\begin{document}

\maketitle

\begin{abstract}
  A significant fraction of stars in globular clusters (about 70\%-85\%)
  exhibit peculiar chemical patterns with strong abundance variations in
  light elements along with constant abundances in heavy elements. These
  abundance anomalies can be created in the H-burning core of a first
  generation of fast rotating massive stars and the corresponding elements
  are convoyed to the stellar surface thanks to rotational induced mixing.
  If the rotation of the stars is fast enough this matter is ejected at low
  velocity through a mechanical wind at the equator. It then pollutes the
  ISM from which a second generation of chemically anomalous stars can be
  formed. The proportion of anomalous to normal star observed today depends
  on at least two quantities : (1) the number of polluter stars; (2) the
  dynamical history of the cluster which may lose during its lifetime first
  and second generation stars in different proportions.  Here we estimate
  these proportions based on dynamical models for globular clusters.  When
  internal dynamical evolution and dissolution due to tidal forces are
  accounted for, starting from an initial fraction of anomalous stars of
  10\% produces a present day fraction of about 25\%, still too small with
  respect to the observed 70-85\%.  In case gas expulsion by supernovae is
  accounted for, much higher fraction is expected to be produced.  In this
  paper we also address the question of the evolution of the second
  generation stars that are He-rich, and deduce consequences for the age
  determination of globular clusters.  
\end{abstract}

\firstsection 
\section{Introduction}

\begin{figure}
  \centering
  \includegraphics[width=0.6\textwidth]{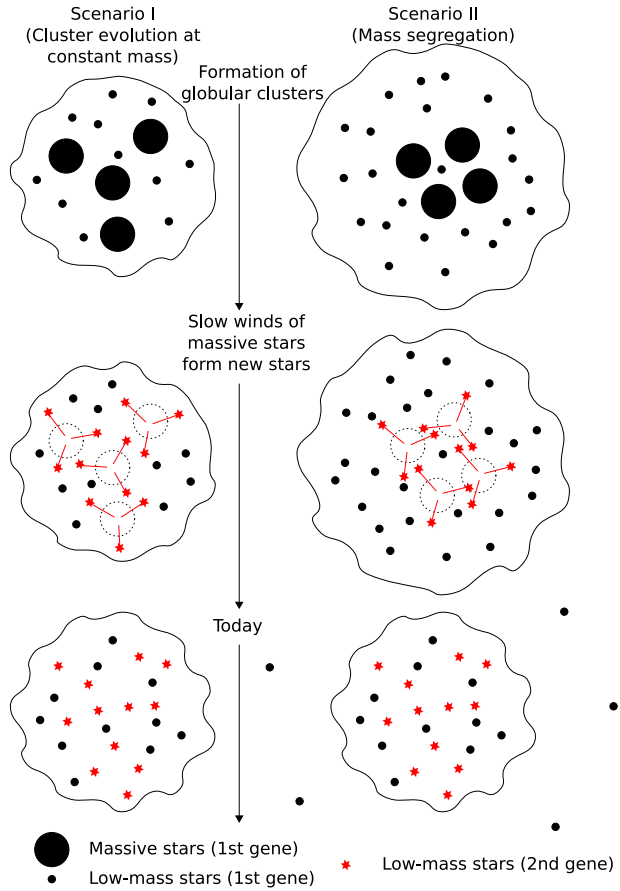}
  \caption{Schematic evolution of a globular cluster: a first generation of
    stars is born from a giant molecular cloud. Massive stars of the first
    generation evolve and give birth to a second generation of low-mass
    stars (dashed symbols in middle panels). Then the cluster evolves and
    today (lower panels) a mixture of first and second generation 
    low-mass stars is present. In the right panel the cluster is initially mass
    segregated and massive stars (and hence stars of second generation) are
    concentrated toward the cluster centre.}
  \label{fig:selfenrich}
\end{figure}

It has long been known that globular cluster stars present some striking
anomalies in their content in light elements whereas their heavy
elements (i.e., Fe-group, $\alpha$-elements) remain fairly constant from star
to star (with the notable exception of $\omega$~Cen).  While in all the
Galactic globular clusters studied so far one finds ``normal'' stars with
detailed chemical compositions similar to those of field stars of same
metallicity (i.e., same [Fe/H]), one also observes numerous ``anomalous''
main sequence and red giant stars that are simultaneously deficient (to
various degrees) in C, O, and Mg, as well as enriched in N, Na, and Al (for 
reviews see \citealp{GrattonSneden2004,Charbonnel2005}).

These abundance variations are expected to result from H-burning
nucleosynthesis at high temperatures around $75\times 10^6$~K
\citep{DenisenkovDenisenkova1989,DenisenkovDenisenkova1990,LangerHoffman1995,PrantzosCharbonnel2007}.
Such temperatures are not reached in the low-mass main sequence and RGB
stars that are chemically peculiar, meaning that the stars inherited their
abundance anomalies at stellar birth.

Here we follow the work of \citet{PrantzosCharbonnel2006} and \citet{DecressinMeynet2007} 
who propose that abundance anomalies are build up
by fast rotating, fast evolving massive stars. 
During their main sequence evolution, rotationally-induced
mixing transports elements synthesised in the convective H-burning core to the
stellar surface. For stars heavier than 20~\Ms{}, the surface reaches break-up 
at the equator (i.e., the centrifugal equatorial force balances gravity), 
providing their initial rotational velocity is high enough.
In such a situation, a slow mechanical wind develops at
the equator and forms a disc around the stars similar to what happens to Be
stars \citep{TownsendOwocki2004,EkstromMeynet2008}.
Matter in discs is strongly enriched in H-burning products and has a slow
escape velocity that allows it to stay in the potential well of the cluster.
On the contrary, matter released later through radiativelly-driven winds
during most of the He-burning phase and then through SN explosion 
has a very high velocity and is lost by the cluster. Therefore, new stars
can form only  
from the matter available in discs with the abundance patterns we observe
today.  
Thus globular clusters can contain two 
populations of low-mass stars: a first generation which has the chemical
composition of the material out of which the cluster formed (similar to
field stars with similar metallicity); and a second generation of stars
harbouring the abundance anomalies born from the ejecta of fast rotating
massive stars. This scenario is sketched in Fig.~\ref{fig:selfenrich}.

\section{Dynamical issues}

\subsection{Number ratio between two populations in globular clusters}

Based on the determination of the composition of giant stars in NGC~2808 by
\citet{CarrettaBragaglia2006}, \citet{PrantzosCharbonnel2006} determined
that around 70\% of stars present abundance anomalies in this specific
cluster today.
\citet{DecressinCharbnnel2007} find similar results for NGC~6752 with their
analysis of the data of \citet{CarrettaBragaglia2007}: around 85\% of the
cluster stars (of the sample of 120 stars) present abundance
anomalies. Therefore most stars still evolving in globular clusters
seem to be second generation stars.

How to produce such a high fraction of chemically peculiar stars? The main
problem is that assuming a \citet{Salpeter1955} IMF for the polluters, the
accumulated mass of the slow winds ejected by the fast rotating massive
stars would only provide 10\% of the total number of low-mass stars. To
match the observations thus requires either (a) a flat IMF with a
slope of 0.55 instead of 1.35 (Salpeter's value), or (b) that 95\% of the
first generation stars have escaped the cluster
\citep{DecressinCharbnnel2007}. Here we first verify whether such a high
loss of stars is possible, and which are the main processes that could drive it.

\subsection{Dynamical evolution of globular clusters}

\begin{figure}[ht]
  \centering
  \includegraphics[width=0.48\textwidth]{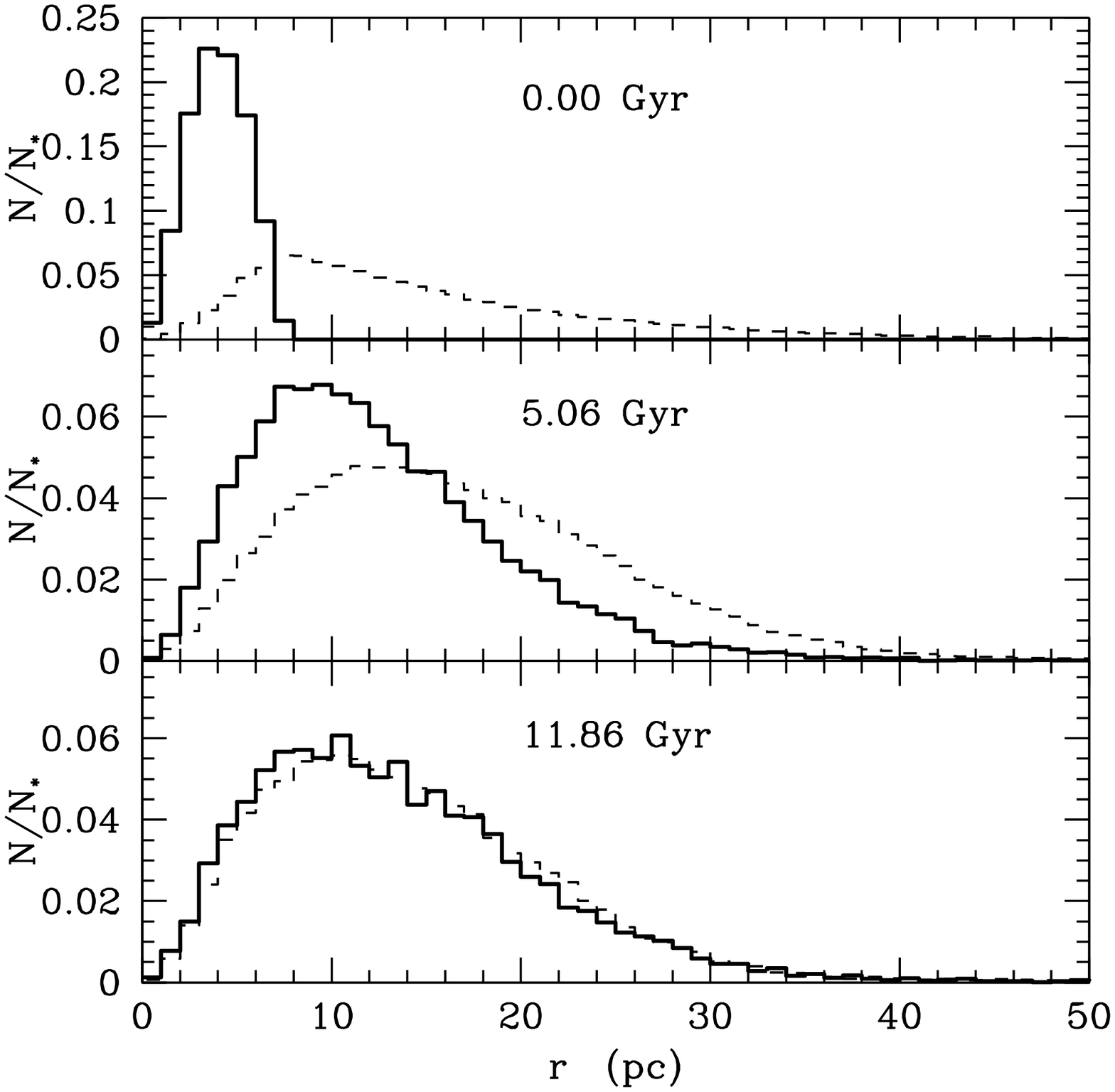}
  \includegraphics[width=0.48\textwidth]{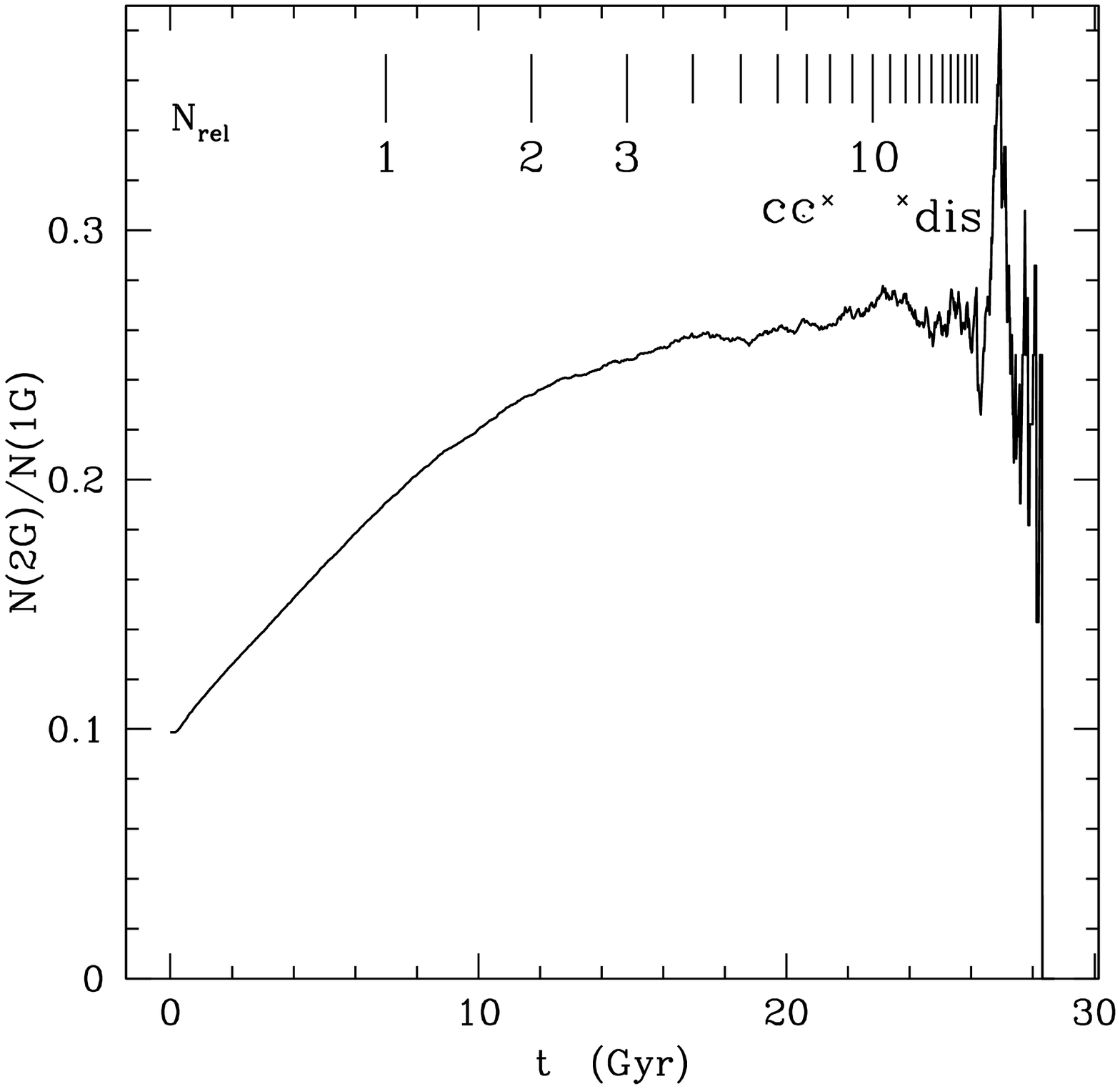}
  \caption{\textit{Left:} radial distribution of the first (dashed lines) and second
    (full lines) generation of low-mass
    stars at three different times. Each histogram is
    normalised to the total number of stars in each population.
    \textit{Right:} Number ratio between the second (with low initial specific
    energy) and first (with high initial specific energy) population of
    low-mass stars in a cluster with initially 128k stars as a function of
    time. At the top of each panel the number of passed relaxation times is
    shown, crosses indicate the time of core
      collapse and of cluster dissolution.
}
  \label{fig:distr}
\end{figure}

First we assume that the globular clusters display primordial mass
segregation so that the massive stars are located at their center.
Since we expect that the formation of the second generation of
low-mass stars happens locally around individual massive stars (see
\citealp{DecressinCharbnnel2007} for more details), the second generation
of stars will also be initially more centrally concentrated than the first
generation. In such a situation, two competitive processes act in the
clusters: the loss of stars in the outer cluster parts will first reduce
the number of bound first generation stars; and the dynamical spread of the
initially more concentrated second generation stars will stop this
differential loss when the two populations are dynamically mixed.

Our analysis, based on the N-body models computed by
\citet{BaumgardtMakino2003} with the collisional Aarseth N-body code
\textsc{nbody4} \citep{Aarseth1999}, is presented in detail in
\citet{DecressinBaumgardt2008}.

As these models have been computed for a single stellar population, we
apply the following process to mimic the formation of a cluster with two
dynamically distinct populations: we sort all the low-mass stars ($M\le
0.9$~\Ms) according to their specific energy (i.e., their energy per unit
mass). We define the second stellar generation as the stars with lowest
specific energies, (i.e., those which are most tightly bound to the cluster
due to their small central distance and low velocity). The number of second
generation stars is given by having their total number representing 10\% of
the total number of low-mass stars.

In Fig.~\ref{fig:distr} one can see the radial distribution of the two
populations at various epochs. Initially, first generation stars show an
extended distribution up to 40~pc whereas the second generation stars (with
low specific energy) are concentrated within 6~pc around the centre.

Progressively the second generation stars spread out due to dynamical
encounters so their radial distribution extends. However this process
operates on long timescale: even after 5~Gyr of
evolution the two populations still have different distributions. 
The bottom panel of Fig.~\ref{fig:distr} shows that after
nearly 12~Gyr of evolution (slightly less than 3 passed relaxation times) the
two populations have similar radial distributions and can no longer be
distinguished owing to their dynamical properties. 

As previously seen, the effect of the external potential of the Galaxy on
the cluster is to strip away stars lying in the outer part of the cluster.
Initially, as only stars of the first generation populate the outer part of
the cluster owing to their high specific energy, only these first
generation stars are lost in the early times. This lasts until the second
generation stars migrate towards the outer part of the cluster. Depending on
the cluster mass, it takes between 1 to 4~Gyr to start losing second
generation stars.  Due to the time-delay to lose second generation stars,
their remaining fraction in the cluster is always higher than that of the
first generation stars except during the final stage of cluster
dissolution.  Fig.~\ref{fig:distr} (right panel) quantifies this point by
showing the time evolution of the number ratio of second to first
generation stars. As a direct consequence of our selection procedure, the
initial ratio is 0.1; it then increases gradually with time and it tends
to stay nearly constant as soon as the two distributions are
similar. Finally, at the time of cluster dissolution (i.e., when the
cluster has lost 95\% of its initial mass, indicated by the label ``dis''
in Fig.~\ref{fig:distr}), large variations occur due to the low number of
low-mass stars present in the cluster. In Fig.\ref{fig:distr} (right panel)
we have also indicated the number of passed relaxation times, showing that
the increase of the number ratio lasts only 3 relaxation times.

The fraction of second generation stars relative
to first generation ones increases by a factor of 2.5 over the cluster
history. Therefore, these second generations stars can account for 
25\% of the low-mass stars present in the clusters. Compared to the observed
ratios (70\% and 85\% in NGC~2808 and NGC~6752 respectively)
the internal
dynamical evolution and the dissolution due to the tidal forces of the host
Galaxy are not efficient enough. An
additional mechanism is thus needed to expel the first generation stars
more effectively.

\subsection{Gas expulsion}

\begin{figure}
  \centering
  \includegraphics[width=0.48\textwidth]{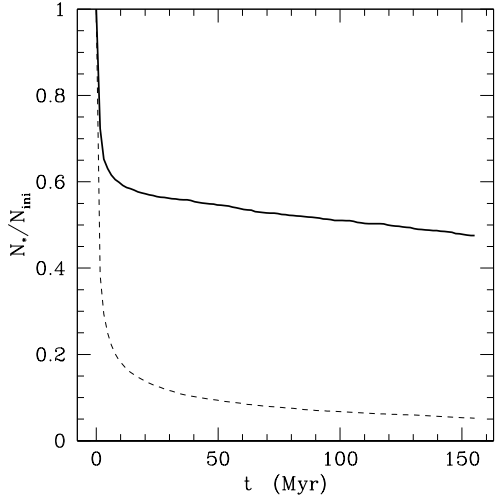}
  \caption{Evolution of the fraction of first (dashed line) and second (full
    line) generation stars still bound to the cluster with
  initial parameters: SFE of 0.33, $r_h/r_t = 0.06$ and
  $\tau_M/t_\text{Cross} = 0.33$.} 
  \label{fig:GE}
\end{figure}

As it operates early in the cluster history (a few million years after
cluster formation at the latest), initial gas expulsion by supernovae is an
ideal candidate for such a process. As the gas still present after the star
formation is quickly removed, it ensues a strong lowering of the potential
well of the cluster so that the outer parts of the cluster can become
unbound.

\citet{BaumgardtKroupa2007} computed a 
grid of N-body models to
study this process and its influence on cluster evolution
by varying the free parameters: star formation efficiency, SFE, ratio
between the half-mass and tidal radius, $r_h/r_t$, and the ratio between
the timescale for gas expulsion to the crossing time,
$\tau_M/t_\text{Cross}$.  They show in particular that, in some extreme
cases, the complete disruption of the cluster can be induced by gas
expulsion.  This process has also been used
successfully by \citet{MarksKroupa2008} to explain the challenging
correlation between the central concentration and the mass function of
globular clusters as found by \citet{DeMarchiParesce2007}.
 
We have applied the same method as the one we used in \S~2.2 to the models of
\citet{BaumgardtKroupa2007}. Fig.~\ref{fig:GE} shows that in the case of
a cluster which loses around 90\% of its stars, the ejection of stars from
the cluster mostly concerns first generation ones. At the end of the
computation only 5\% of first generation stars remain bound to the cluster
along with around half of second generation stars. Therefore the number
ratio between the second to first generation stars increases by a factor of
10: half of the population of low-mass stars still populating the cluster
are second generation stars. Besides, the initial radial distribution is not
totally erased by this mechanism as the second generation stars are still
more centrally distributed. We can expect that this ratio will
continue to increase in the long-term evolution of the cluster (see
Decressin et al., in preparation). 

Thus if globular clusters are born mass
segregated, dynamical processes (gas expulsion, tidal stripping and
two-body relaxation) can explain the number fraction of second
generation stars with abundance anomalies. Similar conclusions have been
reached by \citet{DErcoleVesperini2008}.

\section{He-rich stars}

As abundance variations in light elements are expected to be due to H-burning
whose direct product is He, we expect that second generation stars are
also enriched in He to some degree. Unfortunately He abundance cannot be
directly measured in globular cluster stars and we have only indirect
evidence for an overabundance in He. The globular clusters $\omega$~Cen and
NGC~2808 display multiple main sequences
\citep{PiottoVillanova2005,PiottoBedin2007}; a double sub-giant branch is
also found in NGC~1851 \citep{MiloneBedin2008}. Such features can 
be understood if the stars present various He contents. He enrichment is also
a possible explanation for the appearance of extreme horizontal branches
seen for several globular clusters \citep{CaloiD'Antona2005,CaloiD'Antona2006}.

\subsection{Evolution of He-rich stars}

\begin{figure}
  \centering
  \includegraphics[width=0.48\textwidth]{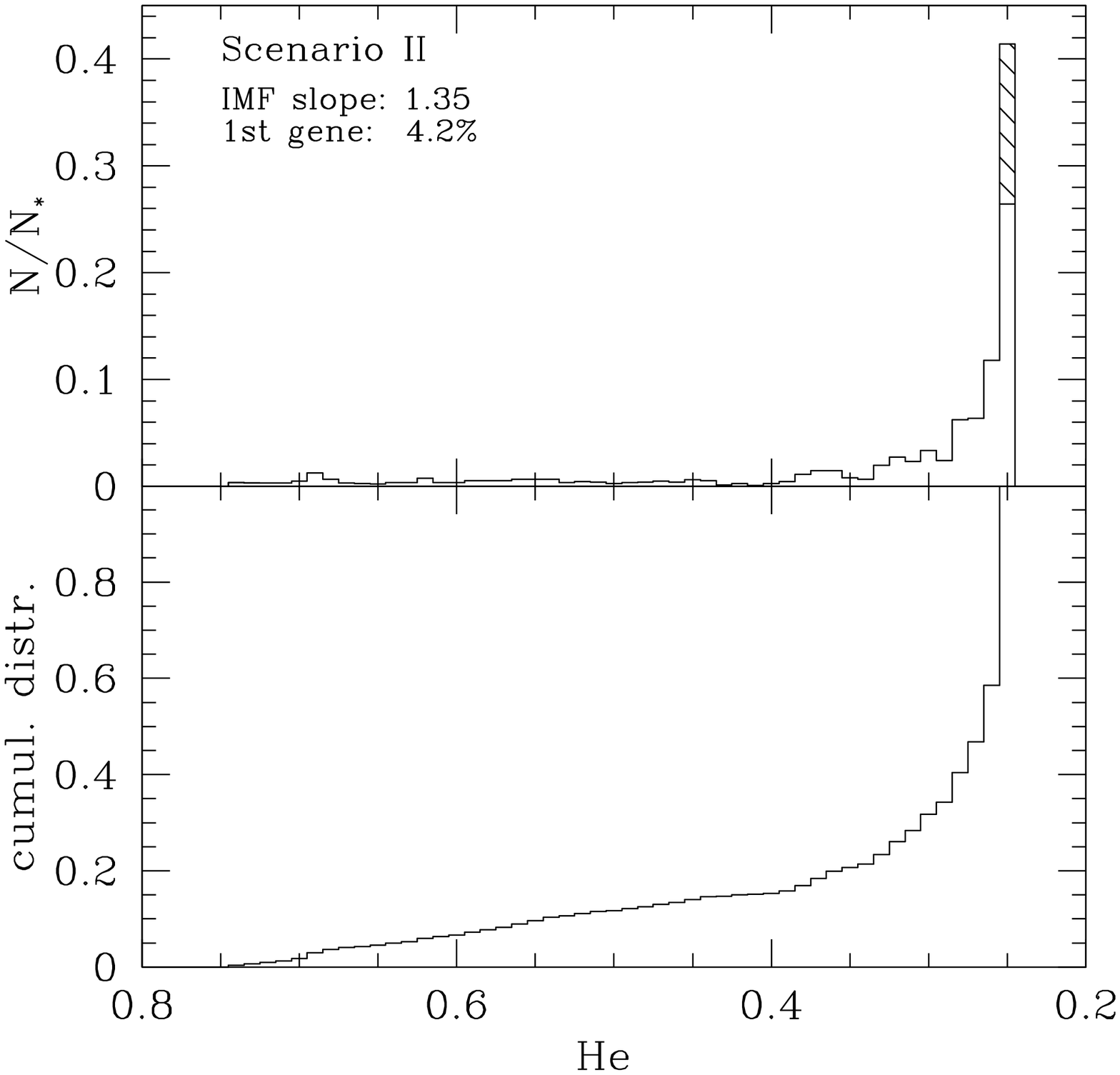}
  \includegraphics[width=0.48\textwidth]{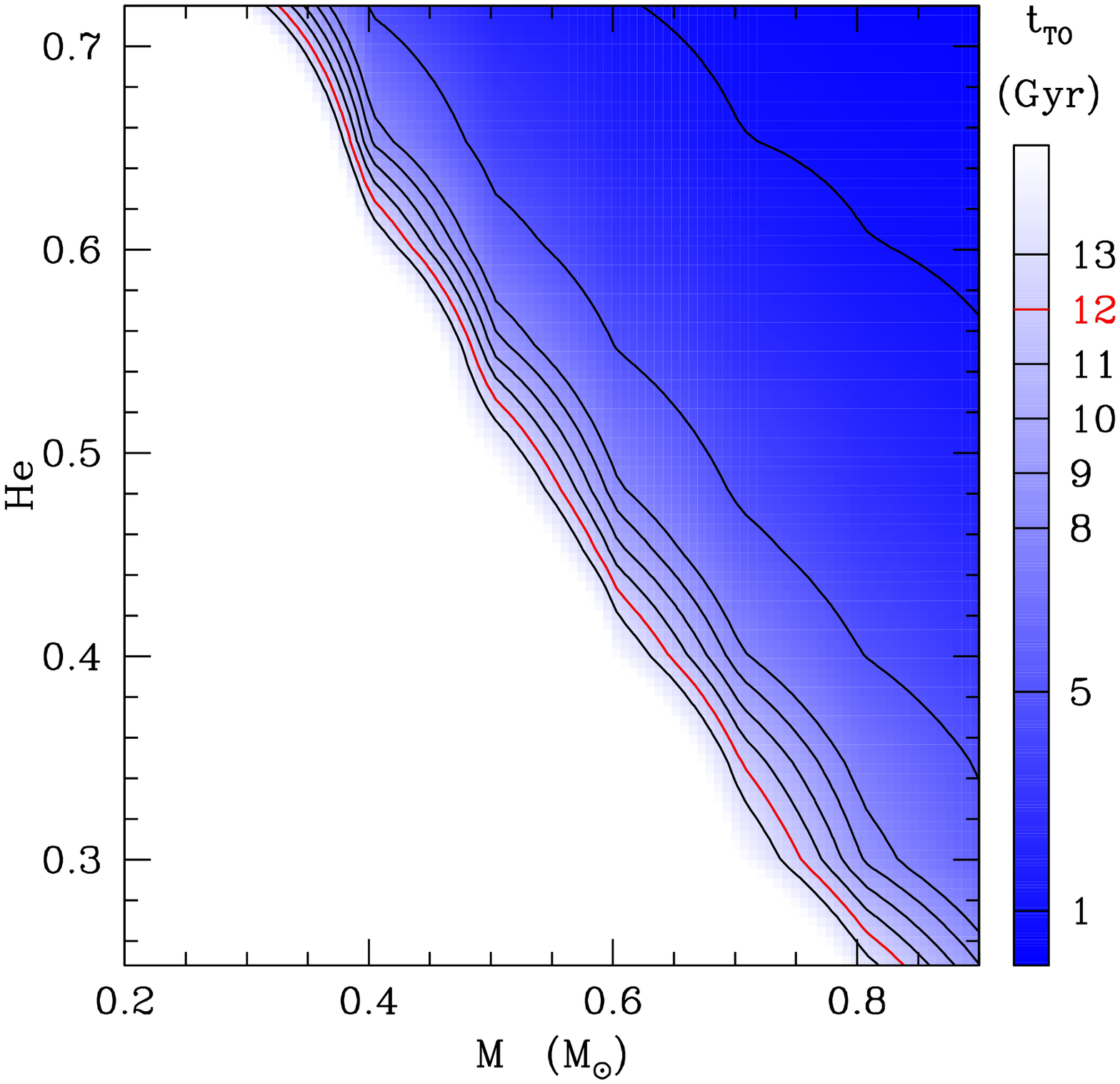}
  \caption{\textit{Left:} distribution function of He for low-mass stars of
  second (white area) and first (hatched area) (top panel) and the cumulative
  distribution function of He in low mass-stars (bottom
  panel). \textit{Right:} Age at the turn-off for low-mass stars as a function
of their mass (0.2--0.9~\Ms) and initial He value (0.245-0.72, mass
fraction). White area indicates stars still on the main sequence after
15~Gyr of evolution.}
  \label{fig:Herich}
\end{figure}

As explained in \citet{DecressinCharbnnel2007}, matter stored in the discs
around massive stars is heavily enriched with He. Fig.~\ref{fig:Herich} (left
panel) gives our expected theoretical distribution function of the He-value
in low-mass stars in NGC~6752. A main peak is present at $Y=0.245$ and
it extends up to 0.4. However a long tail toward higher Y-values is also
present with around 12\% of the stars with initial He value between 0.4 and
0.72.

To assert the implications for globular clusters induced by this population of
He-rich stars we have computed a grid of low-mass stellar models from 0.2
to 0.9~\Ms{} at a metallicity of $Z=0.0005$ (similar to the metal-poor
globular cluster NGC 6752) for initial He mass fraction between 0.245 and
0.72 with the stellar evolution code \textsf{STAREVOL} V2.92
\citep[see][for more details]{SiessDufour2000,Siess2006}. These models have
been computed without any kind of mixing except for an instantaneous mixing in
convection zones. The adopted mass-loss rate follows \citet{Reimers1975}
prescription (with a parameter $\eta_\text{R} = 0.5$) with a 
$\sqrt{Z/Z_\odot}$ dependence. All models have been computed from the
pre-main sequence
to the end of the central He-burning phase.

\begin{figure}
  \centering
  \includegraphics[width=0.48\textwidth]{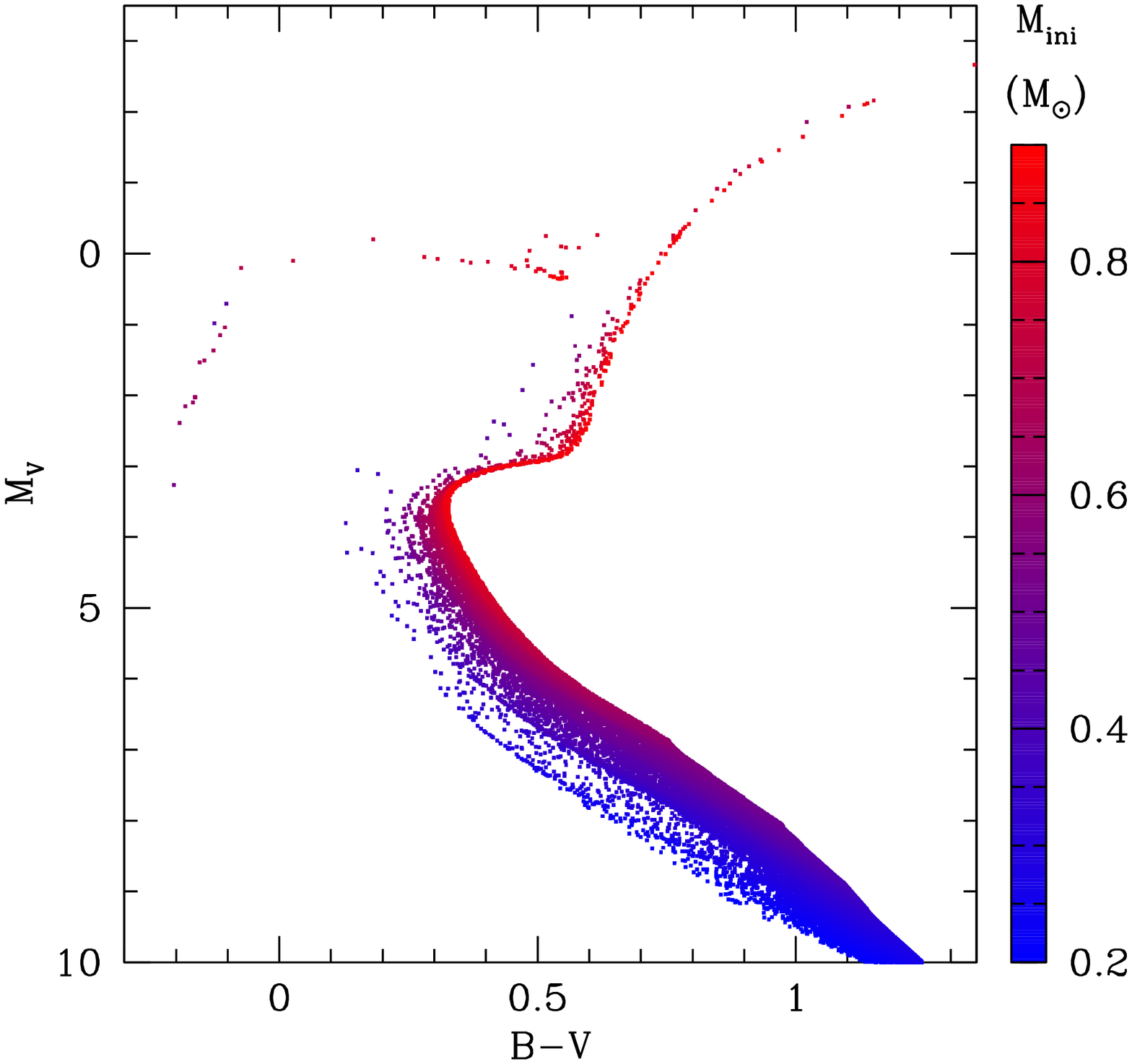}
  \includegraphics[width=0.48\textwidth]{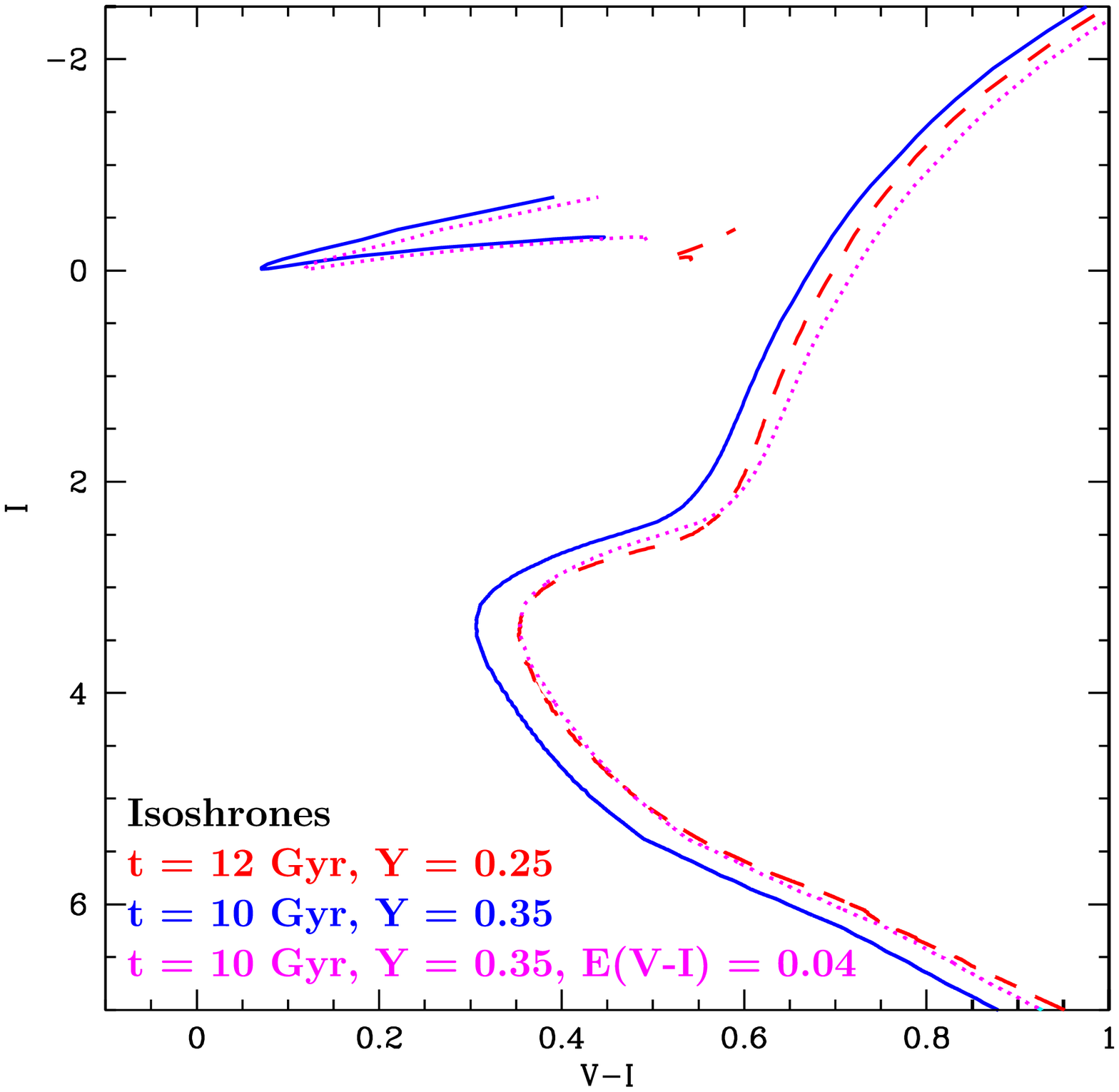}
  \caption{\textit{Left:} Colour-magnitude diagram of a 12~Gyr old globular
      clusters with initial spread in He in stars similar to
      Fig.~\ref{fig:Herich} (left panel). \textit{Right:} isochrone of 12
      Gyr and initial He of 0.25 (dashed lines) and isochrones of 10 Gyr and initial He
      of 0.35 without (full lines) and with (dotted lines) additional reddening.}
  \label{fig:cmd}
\end{figure}

For a given stellar mass, He-rich stars evolve faster on the main-sequence
due to their lower initial H-content and to their higher luminosity. Figure
\ref{fig:Herich} illustrates this point showing the turn-off age as a
function of the initial mass and He mass fraction of stars. After 12~Gyr 
stars of 0.85~\Ms{} with standard helium ($Y = 0.245$) as well as
He-rich stars of 0.4~\Ms{} ($Y=0.6$) are leaving the main sequence.

\subsection{Effects on globular clusters}

Figure \ref{fig:cmd} (left panel) shows a synthetic colour-magnitude
diagram (CMD) of a 12~Gyr old globular cluster with the initial spread in
He as given by Fig.~\ref{fig:Herich} (left panel). This CMD has been
computed with a modified version of the program used by \citet{Meynet1993}
to investigate supergiant populations. The spread in He converts into a
spread in mass at the turn-off.  The luminosity increase of He-rich stars is
mainly compensated by their shorter lifetime so that the turn-off
luminosity is almost constant.  Besides due to their differences in opacity
and to their compactness they are also hotter. Thus the He-rich main-sequence
and RGB stars are shifted to the left side of the CMD. They also occupy
the blue part of the HB down to the extreme-HB location.

If we compare our theoretical CMD with the one observed by \citet{BrownFerguson2005}
for NGC~6752 we note some discrepancies. First the theoretical
width of the main sequence at the turn-off is too large for this
cluster. Additionally NGC~6752 shows on extended blue HB with no stars in the red
part. This last discrepancy could be attributed to the low mass-loss rate used
in the stellar models which do not remove enough mass along the RGB and
hence produce too cool stars on the HB. As the theoretical spread
of the initial He is strongly affected by the dilution of the disc ejected
by fast rotating massive stars and the ISM, we plan to constrain this
dilution  with
the observed sequences to check whether we are able to
consistently reproduce both the abundance anomalies and the He-value
inferred in globular clusters (see Decressin et al., in preparation).

The uncertainties pertaining the ages of globular clusters are manifold.
Among them, photometric
uncertainties widen both sides of the main-sequence, unresolved binaries extend
main-sequence 
towards cooler effective temperature, an increase of metallicity (as seen
in $\omega$~Cen) induces redder sequences. 
The presence of He-rich stars can induce some
additional uncertainties. Let us note that they are the only physical parameter which
broadens the main-sequence only to its left (i.e., blue) side. In
Fig.~\ref{fig:cmd} (right panel) we evaluate uncertainties related to
He-rich stars: we try to reproduce a He-normal isochrone ($Y=0.25$) with a
He-rich ($Y=0.35$) one. This could be done with an isochrone 2~Gyr younger
along with an increase of the reddening of 0.04 magnitudes. 
The differences
between the He-rich with reddening and He-normal isochrones are small
around the TO, the main-sequence and the subgiant-branch. Discrepancies
appear along the RGB and at the level
of the HB, where the normal He-rich isochrone is much less extended toward
the blue. Thus the age uncertainties due to a population of
He-rich stars can be of the order of 1--2~Gyr.

\acknowledgements

T. D. and C. C. acknowledges financial support from the swiss FNS.




\end{document}